\documentclass[journal]{IEEEtran}

\usepackage{graphicx, epstopdf}
\usepackage{setspace}
\usepackage{acronym}
\usepackage{amssymb}
\usepackage[cmex10]{amsmath}
\usepackage{commath}
\usepackage{mathrsfs}
\usepackage{mathtools}
\usepackage{ifthen}
\usepackage{textcomp}
\usepackage{float}
\usepackage{subfigure}
\usepackage{cite}
\usepackage{bm}
\usepackage{url}
\usepackage{siunitx}
\usepackage{booktabs}

\IEEEoverridecommandlockouts

\begin{document}

\title{Urban Analytics: Multiplexed and Dynamic Community Networks}

\author{Weisi Guo\textsuperscript{1,2}, Guillem Mosquera Do\~nate\textsuperscript{3}, Stephen Law\textsuperscript{1,5}, Samuel Johnson\textsuperscript{3}, Maria Liakata\textsuperscript{1,4}, Alan Wilson\textsuperscript{1,6}

\thanks{\textsuperscript{1}The Alan Turing Institute, UK. \{\textsuperscript{2}School of Engineering, \textsuperscript{3}Mathematics Institute, \textsuperscript{4}Department of Computer Science,\} University of Warwick, UK. \{\textsuperscript{5}Bartlett School of Architecture, \textsuperscript{6} Bartlett Centre for Advanced Spatial Analysis,\} University College London (UCL), UK. This work was supported by The Alan Turing Institute under the EPSRC grant EP/N510129/1. Corresponding Author: wguo@turing.ac.uk}}

\markboth{preprint, 2018}
{Submitted paper}

%%%% TOTAL WORD COUNT LIMIT: 4500, 6 figures, 15 references

\maketitle

\begin{abstract}
In the past decade, cities have experienced rapid growth, expansion, and changes in their community structure. Many aspects of critical urban infrastructure are closely coupled with the human communities that they serve. Urban communities are composed of a multiplex of overlapping factors which can be distinguished into cultural, religious, social-economic, political, and geographical layers. In this paper, we review how increasingly available heterogeneous mobile big data sets can be leveraged to detect the community interaction structure using natural language processing and machine learning techniques. A number of community layer and interaction detection algorithms are then reviewed, with a particular focus on robustness, stability, and causality of evolving communities. The better understanding of the structural dynamics and multiplexed relationships can provide useful information to inform both urban planning policies and shape the design of socially coupled urban infrastructure systems.
\end{abstract}


\section{Introduction}

\subsection{Socially Coupled Networks}
Cities are embedded with a multitude of critical infrastructure (CI) systems, including but not limited to wireless networks, transport networks, water distribution networks,...etc. All of these CI systems are closely coupled with the human communities that they serve. Over the past decade, rapid urbanization has changed the traditional community structure within cities. Rapid proliferation of new digital technologies such as social media have overlapped spatially-embedded geographical communities with new online-social-networks (OSNs) interactions. Developing continuous understanding of the evolving urban community structure at different topical layers is important to inform urban policies and the operation of CI systems. As such, we need to understand not only how the community structure varies across different geographical, social, and technological contexts, but also develop robust algorithms to identify the strength of overlapping and interdependent communities. 

In recent years, there is emerging research in using better understanding of social community network structure to drive CI operations. For example, a social device-to-device communication network \cite{Orsino16, Li14} coupled with a social peer-to-peer (P2P) network \cite{Wu16} is heavily dependent on the social interaction data, especially in densely populated urban environments. The close coupling between technology and society means that the CI systems often conform to natural laws, which in turn leads to deeper scientific insight into their topological properties and offers new scalable modeling approaches. For example, the wireless mobile network nodes have been shown to be random uniformly distributed, giving rise to stochastic geometry modeling approaches. These in turn lead to insightful network capacity scaling laws that can inform long-term investment policies \cite{Heath16}. 

\subsection{Multiplexed Community Structures}
Graph theoretic and network science research in the past 2 decades have given rise to strong understanding of singleplex complex networks. However, it is only in the last few years that we are understanding the structure and dynamics of multiplexed networks. Challenges in understanding these multi-layer networks include the following: i) how do we uncover the interdependencies between layers, ii) what is the impact of the interdependencies, and iii) what is the community structure in multiplexed networks.

\begin{figure*}[t]
	\centering
	\includegraphics[width=0.95\linewidth]{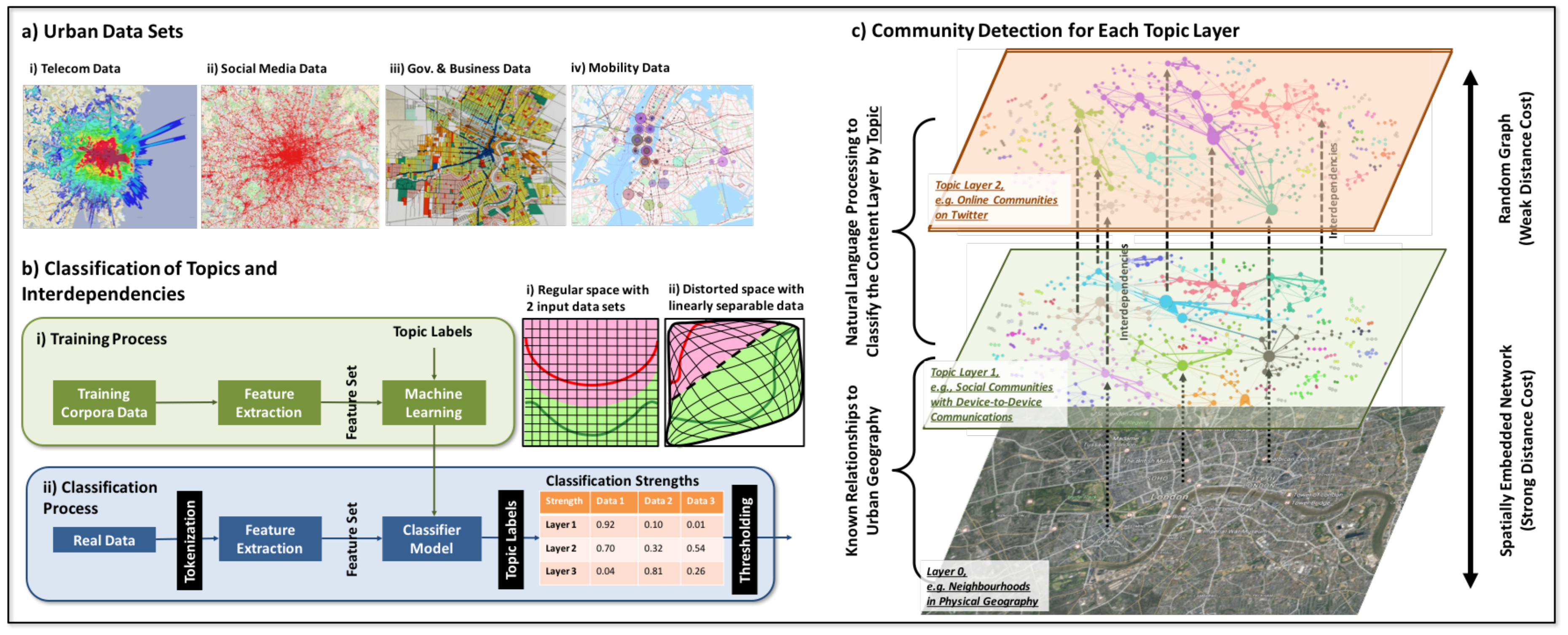}
	\caption{Urban data analytics: (a) Urban data sets, (b) Machine learning classification of topic layers, (c) Communities with interdependencies and different levels of spatial embedding.}
	\label{fig:1}
\end{figure*}

In this review paper, we will examine two methodological developments in community analysis in urban environments. In the first part of the paper, we will review how developments in natural-language-processing (NLP) can help uncover and classify the different aspects of overlapping urban layers (see Fig.~\ref{fig:1}). In the second part of the paper, we will review different community structure detection methods, with a focus on community detection robustness and stability to perturbations. Methodologically, the paper covers a range of useful algorithms from machine learning, statistical physics, and graph theory. In the third part of the paper, to show application and inspire further research, we comment on the relevance of this knowledge to both urban planning and socially-driven communication networks. Finally, we conclude and comment on open challenges in this area.

\section{Uncovering Communities using NLP}\label{topics}
% Maria
Urban communities are an overlap of different topic layers. NLP allows us to study and analyse textual content from online interactions to uncover aspects of communities (see Fig.~\ref{fig:1}). The latter include: (a) discussion topics of common interest (b) public mood and opinions as well as attitudes towards specific entities and issues (c) detection  of user characteristics such as gender or income. The content comes primarily from online social networks (OSNs), including Facebook, Twitter, which is often tagged with location information, as well as  online fora and mobile phone networks. 

\subsection{Probabilistic Topic Modeling}
The most widely used technique for detecting topics of interest is probabilistic topic modeling. This is an unsupervised method which assumes that a collection of data are permeated by hidden underlying topics, each of which is uncovered as a set of \textit{word distributions} that define the topics. There are numerous variants on probabilistic topic modeling to suit different data lengths, but the core of most is \textit{Latent Dirichlet Allocation (LDA)}, where a prior Dirichlet distribution of topics and an iterative generative process is used to discover the conditional distribution of topics given the data and the conditional distribution of words to topics. The iterative step involves computing the above two probabilities given the previous assignment to topics and then re-assign each word to a topic using the probabilities, assuming that each word is being generated by its topic. Recent work has shown that LDA can reveal similar topics representing a set of documents as traditional social science survey methodologies \cite{baumer2017comparing}.

%The goal in opinion mining including, sentiment analysis and stance detection is to discover opinions about topics or issues mentioned in documents or conversations as two or three-way supervised classification problem. While originally sentiment analysis was developed for product reviews and involves classification at the document level, (e.g. one sentiment per review or tweet) in recent years it has moved towards aspect based sentiment and target based sentiment analysis where the objective is to obtain the sentiment towards specific entities (e.g. politicians, topics) or aspects (e.g. the safety of a neighborhood) resulting in multiple types of sentiment per document (e.g. per tweet) \cite{wang-EtAl:2017:EACLlong,DBLP:conf/coling/SaeidiBLR16}. 

\subsection{Classification Techniques}
Most systems obtain features from the text either as sequences of words (e.g. $n$-gram probabilistic $(n-1)$-th order Markov model) or more recently as combinations of word embeddings, where each word encountered in the vocabulary is represented by a \textit{word vector} obtained through iterative methods operating on large word co-occurrence matrices. While for some time Support Vector Machines (SVM) have been used as the classifier of choice for sentiment analysis, in recent years these have been replaced by neural network approaches and in particular convolutional neural networks (CNN) and recurrent neural networks, depending on the particular problem. CNNs require no domain expertise and automatically create increasing levels of abstraction to distort the input data space into linearly separable data (see Fig.~\ref{fig:1}b). CNNs are particularly suited to the synthesis of sentence representations from individual words while recurrent neural networks and most specifically Long Short-Term Memory Networks (LSTMs) are usually preferred for sequential classification tasks. In the classification process, the strength of topic interdependencies can also be uncovered. For example, a data set maybe classified to belong to a number of topic layers, with a vector $L$ indicating the classification \textit{strength} or the \textit{accuracy} of classification for each topic layer. A threshold is then used to decide which layers are selected as being inter-related (see Fig.~\ref{fig:1}b-c).

%Recent work has sought to incorporate social network structure as embeddings which are used to help create notions of linguistic homophily that can guide the focus of neural network approaches in sentiment analysis \cite{yang2017sentiment}.
%Another line of relevant work in NLP involves inferring characteristics of users from their posts on social media. Such characteristics include mood, socioeconomic status \cite{Lampos2016}, gender and age \cite{nguyen2013old}. The problem is usually tackled as a supervised classification task with authors in this area using linear and logistic regression models as well as non-linear regression and kernel methods such as Gaussian processes. Challenges involve ethics concerns on appropriateness of data labeling and biases in data sampling.

In summary, NLP and machine learning techniques enables us to classify data into different topic layers and uncover the strength of interdependencies. In the following sections, we will review how communities are detected from relational data (e.g. call records, re-Tweets or discussion threads) and how knowledge of urban communities can shape policies in urban planning to driving real time resource management in social wireless networks. \\

\section{Urban Community Detection Through User Interactions}

While the previous section is concerned with analysis of content, in the rest of the paper we focus on how communities can be detected through the number and type of interactions. Community structures arise from a variety of interactions. In the context of a city, the interactions stem from people communicating with or traveling to other people. This can occur online through social media, digitally via telecommunications, or physically through different transportation mechanisms. Most of the interactions can be classified on a spatial axis, where the friction of the interaction determines whether the community network structure will be a \textit{spatially embedded network} (i.e., walking-distance neighbourhood) or a \textit{random graph} (i.e., online social network) - see Fig.~\ref{fig:1}c. Community structures arise when a group of interactions have a stronger mutual coupling than with other interactions. Increasingly available data, such as mobile call detail records (CDR) can reveal and quantify the strength of community interactions (e.g. detecting community boundaries from CDRs in Belgium \cite{Blondel08}). In this section, we review how communities can be detected, what measures exist for understanding its robustness, and whether the communities are resilient to the rapid changes undergoing in cities.

\subsection{Review of Community Detection Methods}

There are a number of ways to detect and define community structures from the underlying data of interactions between different urban entities presented above in Section II. However, given the ill-defined nature of network communities, selecting a suitable detection method is still discretionary to the researcher's needs and intuition, both in terms of computing complexity and data characteristics. Here we present the main general classes of community detection methods currently in use across the literature, referring to their strengths and weaknesses.

\subsubsection{Spectral Graph Clustering}
A long known result in graph theory is that clusters of highly inter-connected vertices can be recovered through the study of the \textit{eigenvalue spectrum} of the adjacency or Laplacian matrices of the graph \cite{Newman13}. It can be seen that, for most non-sparse networks, one or more eigenvalues of these matrices appear as outliers in the complex plane with respect to the rest. The aim of spectral graph clustering then is to use the eigenvectors corresponding to such outlying eigenvalues to project the vertices to a metric space where they can be easily clustered using unsupervised techniques such as \textit{k-means}. Unfortunately, there's no guarantee of finding convergence in the clustering step for sparse networks, which is typically the case of real-world networks.

\subsubsection{Modularity}
Many community detection methods are based on the optimization of some cluster quality function $Q$ over the space of possible partitions of the network. The most popular quality function was introduced by Girvan and Newman: so-called \textit{modularity} computes, for every cluster in a partition, the difference between the empirical intra-cluster edge density and the corresponding density expected in a chosen null-model \cite{Newman13}. In this sense, modularity is a quality function of the form $Q = \sum_{ij}\big(A_{ij}-P_{ij}\big)\delta(c_i,c_j)$, where $A_{ij}$ is the empirical adjacency matrix, $P_{ij}$ the expected edge weight in the null model, and $\delta(c_i,c_j)=1$ when nodes $i$ and $j$ belong to the same community but vanishes otherwise. The configuration model - i.e. a degree-preserving randomized version of the network - is usually chosen as a null-model. The rationale is that a complete randomization of a network destroys any existent community structure. Therefore, a \textit{modularity-maximizing partition} ensures the presence of meaningful communities. However, such maximization over the partition space has been shown to be \textbf{NP-hard}, and consequently a number of more or less successful heuristics have been developed in order to find approximated solutions - for instance, greedy optimization of modularity in the street network of Greater London using the Louvain method is shown later in Fig.~\ref{fig:4} \cite{Law2016}. Furthermore, modularity optimization suffers from a systematic bias called \textbf{resolution limit}: this effect prevents the method from finding small communities in large networks, effectively making it unreliable for some multi-scale real-world applications.

\subsubsection{Statistical Inference}
Statistical inference is a logical approach towards community detection when both the generative network model of the data and the number of communities present in it are supposed to be known. The empirical network is treated as a sample from the generative model with a given set of parameters, and the aim is to infer such parameters using likelihood maximization techniques. The most popular generative model is the Stochastic Block Model (SBM), where the main parameters are the probabilities of finding edges between nodes inside every community, and between nodes across every different pair of communities \cite{Newman13}. Despite being exact and conceptually simple, the method necessarily needs to sample across the space of possible partitions of the network to find a guaranteed maximum of likelihood, making it \textbf{NP-complete}. Multiple approximations to the problem exist that guarantee high-probability recovery of the community structure. As in the case of modularity, this method suffers from the \textbf{resolution limit}.

\subsubsection{Network Dynamics}
A more recent approach to the study of networks consist in analyzing the properties of dynamical processes running on the graph \cite{Delvenne13}. The assumption here is that the interplay between dynamics and structure can yield a better understanding of the mesoscopic organization of the network. In the case of community detection, three main classes of dynamical processes running on the nodes of the network are usually considered: spin-spin interaction models, synchronization of phase oscillators and diffusion processes. Interestingly, the first two processes can be shown to be dual to the later, i.e. equivalent to Markov chain modelling the diffusion of random walks across the network. Furthermore, using different conditions one can see that clustering nodes based on diffusion flow characteristics is equivalent to optimizing modularity in one extreme and to performing spectral clustering in the other: studying Markov processes on the graph is therefore one of the most versatile ways of unveiling the community structure of a network. These diffusion processes can be easily interpreted in the context of OSN data, insomuch they represent the flow of information, interests and influence across the underlying social interaction network. Finally, random walk diffusion can be used to study the community structure of multiplex networks, a generalization that the aforementioned non-dynamical methods cannot naturally accommodate. Further details on the stability and multiplexing of the Markov method are given below in sections \ref{markov_stability} and \ref{multiplex}.

\begin{table}
\small
\renewcommand{\arraystretch}{1.5}
\centering
\caption{Summary of the main methods for detecting communities in complex networks (references in the text)}
\label{my-label}
\begin{tabular}{cc}
\hline
Detection Method        & Community Indicator     \\ \hline
Spectral Clustering     & Eigenspace Closseness   \\
Modularity Optimization & Higher Link Density     \\
Statistical Inference   & Higher Link Likelihood  \\
Spin-Spin Interactions  & Low Energy Domains      \\
Coupled Oscillators     & Phase Synchronization   \\
Markov Processes        & Random Walk Confinement \\ \hline
\end{tabular}
\end{table}

\subsection{Stability of Communities Across Markov Time-Scales} \label{markov_stability}

Communities can be regarded as bottlenecks for diffusive flows propagating through the link structure. In particular, the robustness of a community can be quantified by its capacity to contain a stationary random walk between its nodes longer than otherwise expected by chance. In that respect, one needs to define the time-horizon (i.e. the number of Markov steps) for which such retention is measured: at short time-scales, smaller communities will appear as sufficiently robust, whereas at longer time windows only increasingly larger communities will be able to contain the flow. 

More formally, the \textit{stability} of a partition $\mathscr{P}$ under a Markov process $\mathcal{M}$ at time-scale $t$ is defined as follows:
\begin{equation}\begin{split}
    \label{markovian_stability}
    R_{\mathcal{M}}(t) = \sum\limits_{c\in \mathscr{P}} \Big[ P(c,t) - P(c,\infty) \Big] .
\end{split}\end{equation}
Here, $P(c,t)$ refers to the probability of finding a stationary random walker starting in community $c$ within that same community after time $t$. Under ergodicity conditions, $P(c,\infty)$ can be interpreted as the probability of finding two independent stationary random walks in $c$, thus performing the role of a null model.
As mentioned above, Markov time $t$ is here regarded as a multi-resolution parameter allowing to zoom in from finer to coarser grained partitions of the network, providing an understanding of the multiscale nature of real-world network.

For random walks on a singleplex, undirected graph represented by the adjacency $A_{ij}$, the continuous-time density of walkers on any node $i$ evolves as $\dot{p}_{i}=\sum_j \frac{A_{ij}}{k_j}p_j-p_i$. Such expression can be generalized for different network structures or dynamics, and stationary density distributions can be derived and used to obtain specific applicable extensions of (\ref{markovian_stability}).

\begin{figure}[t]
	\centering
	\includegraphics[width=0.95\linewidth]{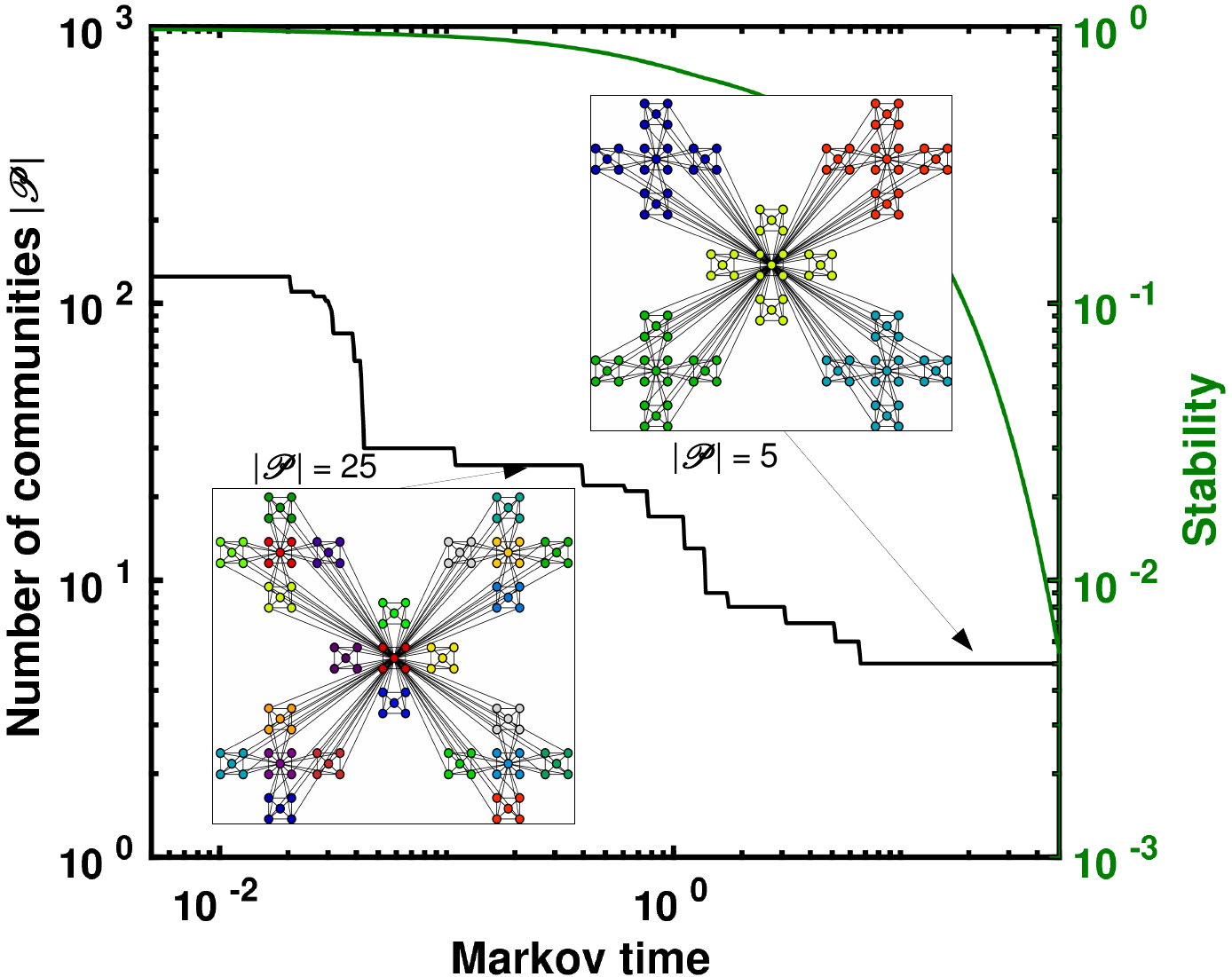}
	\caption{Markov stability (green) and number of communities (black) for the optimal partitions across time-scales for the multi-scale hierarchical network proposed in \cite{Ravasz03}, using the Louvain method. }
	\label{fig:2}
\end{figure}

To summarize, this method is similar to statistical inference or modularity maximization in that it seeks optimization of a quality function, in this case Markov Stability: thus, it can be argued to be computationally \textbf{NP-complete}. However, proven by the exact convergence of stability to modularity at time $t=1$ \cite{Delvenne13}, one can use any of the multiple existing heuristics to sample the space of partitions designed for the latter method - e.g. the \textbf{Louvain algorithm} - obtaining an array of meaningful partitions that can then be classified according to their stability at every Markov time. It is interesting to note how this method allows for a natural definition of community robustness: \textit{partitions deemed most Markov stable through several consecutive time-scales should be understood as the most robust.} Fig.~\ref{fig:2} illustrates this for the hierarchical scale-free graph proposed by Ravasz and Barab\'asi \cite{Ravasz03}: the two natural partitions (with 5 and 25 communities respectively) correspond the longest plateaus of stability across Markov time, i.e. are the most robust. 

\subsection{Communities in Multiplex Networks}\label{multiplex}

Finally - and most important for our urban analytics application scenario - Markov stability methods have been generalized to multiplexed networks, where different singleplexes represent different content layers or the critical infrastructures that serve them. In this case, random walks diffuse inside every slide $s$ of the network following the intra-layer adjacency matrices $A_{ijs}$, but can also jump across layers according to the interdependent connections represented by the tensor $C_{jsr}$, signaling the presence ($C_{jsr}=\omega$) or absence ($C_{jsr}=0$) of a reciprocal connection of node $j$ across layers $s$ and $r$. Note how the strength of inter-layer (interdependence) coupling, modulated by $\omega$, corresponds to the threshold mentioned in Section \ref{topics} for the case of urban modelling. Similarly to Markov time, this parameter could uncover multiple resolutions of the community structure in multiplex urban networks: higher coupling favours partitions that merge different topic layers, whereas low coupling will yield strongly topic-dependent communities. In general, $\omega$ brings a way of studying the overlapping nature of communities in temporal, multiscale or multiplex networks.
\begin{figure}[t]
	\centering
	\includegraphics[width=1.00\linewidth]{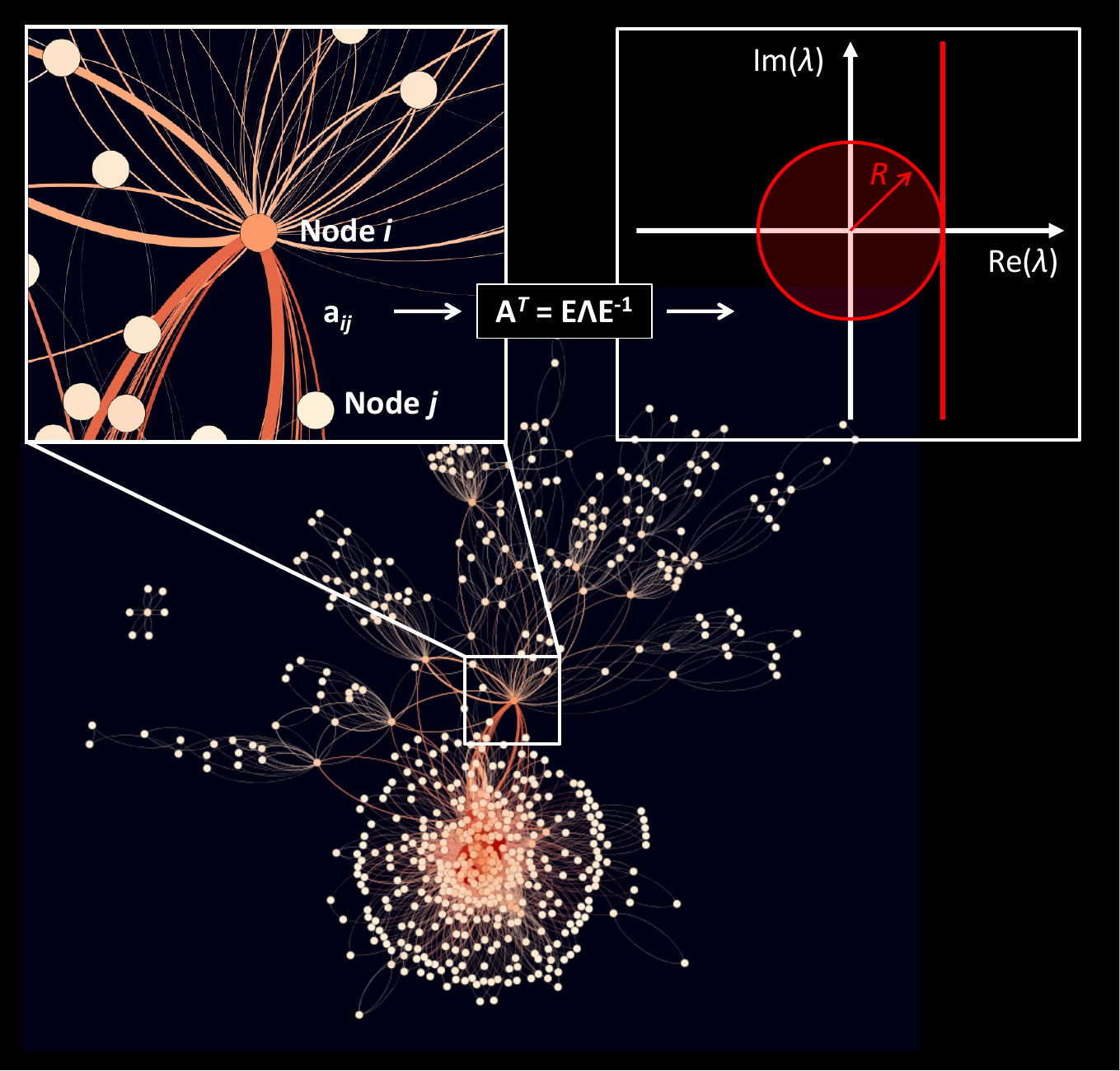}
	\caption{Network stability to perturbations: stability criterion based on leading eigenvalue $R$ of the interaction matrix \textbf{A}.}
	\label{fig:3}
\end{figure}

\subsection{Resilience to Perturbations}

Over the past few years, there has been growing interest on whether large-scale complex networks can be stable as they grow. It has been shown that for any general complex non-linear system where there are coupled dynamic elements governed by linear differential equations, stability can still be inferred. For example, consider a network element with $N$ nodes and the $i$-th node has a property $x_i$ that is of interest. Then changes in its property can be generally written as: $\dif x_{i}(t) /\dif t = -Cx_{i}(t) + \sum_{j}^{N}a_{ij}x_{j}(t)$, whereby $a_{ij}$ is the link strength between nodes and $C$ can be interpreted as some innate capacity inside the node that is resistant to change. For Lyapunov stability, whereby small perturbations do not cause runaway fluctuations in the size of the system, it has been shown that the leading eigenvalue $R$ of the matrix scales with square root of the mean degree of each node $D$: 
\begin{equation}\begin{split}
    \label{stability_scaling}
    R \propto \sqrt{ND} < C.
\end{split}\end{equation} 
As such, even without precise knowledge of the complex dynamics, one can still check the stability of a system from the network structure, e.g., the leading eigenvalue is bounded by $R<C$ (see Fig.~\ref{fig:3}). This gives rise to the well known stability and size trade-off, whereby a system can become unstable if there are too many elements or too many connections. More recent research has shown that by enforcing \textbf{energy hierarchy coherence}, networks can be grown to arbitrary scale and connectedness \cite{Johnson14}.

Such analysis can be extended to the different scales of urban networks. For example, $a_{ij}$ can describe a coarser representation of the network in which nodes now represent communities stemming from a suitable detection algorithms. In the case of a multiplexed structure, understanding the stability of the ecosystem of communities can give rise to more resilient ways to connect people across the different singleplexes of digital (e.g. broadband, online social network), physical (e.g. transport links), and cultural dimensions.

\subsection{Entropy Maximization and Causal Mechanisms}

In order to gain causal insight for observed interaction strengths $a_{ij}$, one can relate the data to known models. Simple non-competitive models (e.g. pairwise gravity law or radiation relationships) have been used for half a century to model economic trade to mobility flow, and a thorough review of gravity laws can be found in \cite{barthelemy2011spatial}. These models provide the underlying benefit and distance cost functions for causal understanding. In terms of benefit functions, almost all research will agree that population determines the flow of goods or people. Where models diverge is in cost functions. For example, using mobile phone connection data, it was found that people had a probability of connecting $\propto d^{-2}$ and for social media the probability was less sensitive to distance, $\propto d^{-1}$. Stepping aside from logical arguments, which state that 2-dimensional entropy maximization leads to a $d^{-1}$ scaling law, there are also distance regimes where the law breaks down, or takes on a Levy flight pattern, e.g. $\propto \exp(-d)$, or becomes insensitive to distance. 

A general class of entropy-maximizing competitive models is known as the \textbf{Boltzmann-Lotka-Volterra (BLV) model}. Using the benefit and cost functions, the BLV model improves significantly over the aforementioned pairwise models. The BLV performs a 2-stage maximization to estimate the weights of the interactions \cite{Wilson08}. The likelihood of a flow $F_{ij}$ is $W(\{ F_{ij} \}) = \frac{F!}{\prod_{ij}F_{ij}!}$. First, it maximizes the likelihood of spatial interaction flow between $N$ entities, which leads to a Shannon capacity form:
\begin{equation}\begin{split}
    \label{entropy_max}
    \max{\log{W(\{ F_{ij} \})}} &= -\sum_{ij}F_{ij}\log{F_{ij}}.
\end{split}\end{equation} After which, we leverage on the aforementioned benefit and cost functions and apply the method of Lagrangian multipliers to take into account the nature of competitive interactions and conservation of flow energy. This method of modeling in urban environments have been shown to accurate for a wide range of applications, including shopping spending behaviour and threat projection between militia parties. Using the data, we can reverse estimate the likelihood of different benefit and cost function parameters to obtain insight into the causal mechanisms behind community structures for different content layers. 

Using longitudinal community interaction data $X^{k}$ and cost-benefit parameters $Y^{k}$ up to time step $k$, one can further check for the \textbf{strength of causal mechanisms} using the directed information: $I(X^{K}\rightarrow Y^{K}) = \sum_{k=1}^{K} I(X^{k};Y^{k}|Y^{k-1})$, where $I(X^{k};Y^{k}|Y^{k-1})$ is the conditional mutual information. \\

\section{Application Domains}

In this section, we review 2 application areas for detecting multiplexed community network structures. First, we review how better understanding of community neighbourhoods can inform urban planning and understand political divisions. In the second part, we review how dynamic community data can be exploited for improving the performance of social telecommunication systems such as social-D2D radio resource management.

\subsection{Urban Planning \& Space Syntax}
\begin{figure}[t]
	\centering
	\includegraphics[width=1.00\linewidth]{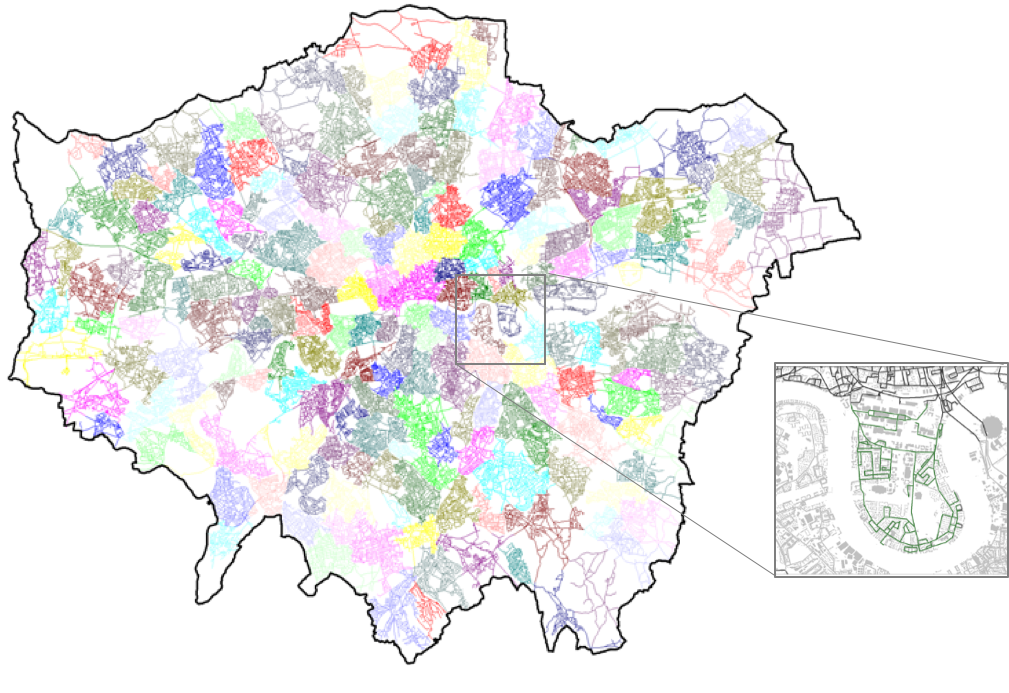}
	\caption{Community Detection on the street network of Greater London reveals a patchwork of neighbourhoods through the city. Highlighted in green is the Isle of Dog in East London. Greedy optimization of modularity using the Louvain Method was applied for this case.}
	\label{fig:4}
\end{figure}

% Stephen 
The use of community detection techniques in urban planning is generally related to the concept of geography or neighbourhood in cities. Neighbourhood can be define as a geographical construct where historical, social, economic, cultural and spatial attributes are homogeneous or connected within, hierarchically nested and dynamic across time. One example of applying network science methods in urban planning is from the field of space syntax, where the street network is used to define a spatial planar graph. Applying the Louvain method of modularity optimisation \cite{Blondel08} on the street network can reveal street-based communities that are cognitively and physically distinct. One such case shown in Fig.~\ref{fig:4} is the Isles of Dogs in East London with few connections to surrounding neighbourhoods \cite{Law2016}. Applying community detection techniques on the commuting flow network can also be used to uncover functional boundaries of city regions \cite{DeMontis2013}. While the adoption of hierarchical percolation theory on the street network can begin to explain the current political divide in the Great Britain.

The importance in understanding the community structure of cities is that it allows urban planners to tackle issues such as social inclusion and economic regeneration through the planning of physical infrastructure. For example, building a bridge between two neighbourhoods can increase the likelihood for more social and economic interactions between them. A case in point is the Millennium Bridge in London, UK where space syntax spatial network analysis were used as part of the design and planning of the pedestrian bridge. Physical infrastructure is seen here as a catalyst for urban regeneration. The use of community detection in urban street network is still in its infancy. The advent of large scale social, cultural, physical and transport network datasets bring with it the opportunity to better understand and potentially model these multi-layer urban challenges.

\subsection{Social Communication Networks}
% Weisi
Social information is also gaining increasing attention in the optimization of wireless RATs. For example, local Device-to-Device (D2D) networks have been shown to conform to social network topologies \cite{Orsino16, Li14}. The underpinning reasons include common social interests and peer discussions that drive common data transfer between entities. These community structures at both the social and digital layers are important to understand and exploit. By harvesting data in a community and utilizing multi-hop D2D communications, recent research has shown that D2D communications can be optimized for local P2P file sharing, leading to improved user experience and cellular traffic offload. For example, the system designed in \cite{Wu16} features cross-layer integration of: 1) a wireless P2P protocol based on the BitTorrent protocol in the application layer; 2) a simple centralized routing mechanism for multi-hop D2D communications; 3) an interference cancellation technique; and 4) a radio resource management (RRM) scheme to mitigate the cross-RAT interference while maximizing the throughput of D2D communications. More recently, several research outputs have proposed optimal RRM using matched physical-social graphs, cooperative game theory based on social group utility functions, and multi-step coalition games of maximize the payoff of cooperating communities \cite{Wang16}. This area of research of coupling telecommunications with social communities is a hot and active area, especially as the line between human and physical systems increasingly blur. \\

\section{Conclusions and Open Challenges}

In the past decade, humanity has experienced both rapid urbanization and proliferation of digital interactions. These processes have rapidly transformed traditional urban landscapes, overlapping spatially-embedded geographical communities with new online social communities. This paper has reviewed methods that can detect community layers and community structures. Using machine learning methods, we can uncover topical layers from data sets. We focus on reviewing and comparing a range of community detection methods, including spectral clustering, modularity optimisation, statistical inference, statistical physics models, and Markov process algorithms. In particular, we discuss measures which can help us understand the robustness, the stability, and the causality of these dynamic and multiplexed communities. We also review entropy maximizing interaction laws that can reveal the underlying causal mechanisms behind the observed data. Open challenges remain in community detection, and we have identified: (1) how to detect overlapping communities using the random walk methodology, and (2) how to uncover the strength  of their interdependencies from a network science perspective.

Finally, we reviewed application domains related to community detection. Many critical infrastructure systems are closely coupled to the social communities that they serve. Developing robust community detection methods allows: (1) urban planners to design better urban spaces through space syntax methods, and (2) network engineers to optimize radio resource application in socially-coupled D2D and P2P communication networks.

In the past few years, there is increased interest in applying machine learning and network science methods to mobile urban data. By reviewing state-of-the-art methodologies and application domains, we hope the paper will inspire the research community to better understand the dynamics of our cities and use the knowledge to inform the design of critical infrastructure systems. \\

\bibliographystyle{IEEEtran}
\bibliography{IEEEabrv,CM_Ref2}

\end{document}